# Statistical Analysis of the Probability of Interaction of Globular Clusters with Each Other and with the Galactic Center on the Cosmological Time Scale According to Gaia DR2 Data

*Maryna Ishchenko[1], Margaryta Sobolenko[1], Peter Berczik[2,1,3] and Taras Panamarev[4,5]*

[1]Main Astronomical Observatory, National Academy of Sciences of Ukraine
[2]Astronomisches Rechen-Institut, Zentrum fur Astronomie, University of Heidelberg, Monchhofstrasse 12-14, 69120, Heidelberg, Germany
[3]Konkoly Observatory, Research Centre for Astronomy and Earth Sciences, Eotvos Lorand Research Network (ELKH), MTA Centre of Excellence, Konkoly Thege Miklos ut 15-17, 1121 Budapest, Hungary
[4]Fesenkov Astrophysical Institute, Almaty, Kazakhstan
[5]Rudolf Peierls Centre for Theoretical Physics, Oxford, United Kingdom
E-mail: marina@mao.kiev.ua

**Abstract**—This study is aimed at investigating the dynamic evolution of the orbits of stellar globular clusters (GCs). To integrate the orbits backward in time, the authors use models of the time-varying potentials derived from cosmological simulations, which are closest to the potential of the Galaxy. This allows for estimating the probability of close passages ("collisions" herein) of GCs with respect to each other and the Galactic center (GalC) in the Galaxy undergoing dynamic changes in the past. To reproduce the dynamics of the Galaxy in time, five of the 54 potentials previously selected from the IllustrisTNG-100 large-scale cosmological database, which are similar in their characteristics (masses and dimensions of the disk and halo) to the current physical parameters of the Milky Way, are used. With these time-varying potentials, we have reproduced the orbital trajectories of 143 GCs 10 Gyr back in time using our original ϕ-GPU high-order N-body parallel dynamic computer code. Each GC was treated as a single physical particle with the assigned position and velocity of the cluster center from the Gaia DR2 observations. For each of the potentials, 1000 initial conditions were generated with randomized initial velocities of GCs within the errors of the observational data. In this study, we consider close passages to be passages with a relative distance of less than 100 pc and a relative speed of less than 250 km/s. Clusters that pass at longer distances and/or with higher velocities do not have a substantial dynamic effect on the orbits of GC. In our opinion, the largest changes in the orbits of clusters can be caused by clusters that pass with low velocities at distances smaller than several fold (for example, fourfold) the sum of the radii of the cluster half-masses. Therefore, the authors regard such close passages separately (for brevity, we will call such passages "collisions"). To select clusters that pass at close distances from the GalC, the following criterion is applied based only on the relative distance: it must be less than 100 pc. Applying the above criteria, the authors obtained statistically significant rates of close passages of GCs with respect to each other and to the GalC. It has been determined that GCs during their evolution have approximately ten intersecting trajectories with each other on the average and approximately three to four close passages near the GalC in 1 Gyr at a distance of 50 pc for each of the chosen potentials.

**Keywords:** numerical methods, Galactic globular clusters, evolution of Galaxy, IllustrisTNG-100, kinematics and dynamics of Galaxy, center of Galaxy

INTRODUCTION

According to the ΛCDM cosmological model, stellar globular clusters (GCs) are the first stellar associations that formed in the early universe and are gravitationally bound star systems older than 10 – 12 Gyr [28] with a current mass of approximately $10^5$ M$_\odot$ on average [17]. These ancient remnants preserve the imprint of the early active era of the formation of our Galaxy [12]. Thus, these objects can be used as a powerful tool for studying the Galactic structure and the history of the formation of the Galaxy itself at various scales starting from the formation of star clusters up to hierarchical events of galactic cannibalism [18]. Taking into account the time that the clusters spend in our Galaxy, we assume that they can pass at close distances with respect to each other during their orbital evolution and could even interact with each other during a collision by exchanging stellar population. On the other hand, we also assume that a certain type of GC orbits should lie near the GalC. It should also be noted that such close passages of the GC near the GalC could cause the loss of the stellar population by the GC, which could partially move to the region of the nuclear star cluster in the GalC. Trying to reproduce – at least approximately – the structure of the Galaxy, we took data from the IllustrisTNG-100 cosmological simulation database and created several time-varying numerical potentials by the approximation method. These cosmological models are among the best ones available today [25]. The dynamic potentials we obtained, such as the current mass and size of the halo and the current mass and size of the disk, have the main parameters that are close to those of our Galaxy [20].

To reproduce the integration of the GC orbit back in time, we used a consolidated catalog that was created using the available catalogs [6, 32]. The combined catalog has 152 objects and contains full 6D spatial–phase information for each object. High-precision data on astrometric measurements, such as the measurements of positions and velocities, from these catalogs in the Gaia Data Release 2 (DR2, [11]) allowed us to reconstruct the trajectories of GC orbits with high precision when integrated down to 10 Gyr back in time.

It should be noted that studies devoted to the kinematics and chemical composition of the GC began to appear in recent years with the appearance of fairly accurate positions and velocities of the GC. There is an example of a study that uses the integration of orbits in time to analyze the dependence of the type of orbit parameters on the percentage of metals in the GC [3]. The authors of [3] argue that most metalrich stellar clusters have orbits with a predominantly low eccentricity, whose angular momenta and orbital planes are similar to those of the Galactic disc.

In [26, 27], the integration of GC orbits in a constant potential was performed and it is established that approximately 30% of the clusters, which were classified as belonging to the bulge based on their positions, actually pass by merely the inner region of the Galaxy. Most likely, they belong to the inner halo or thick disk component. Most of the GCs confirmed as belonging to the bulge do not follow the structure of the bar and are older than the era of the formation of the bar itself.

There are interesting studies [29, 13, 22, 1, 2] devoted to the integration of GCs with components, such as a bar and spiral arms, which are similar to our Galaxy, in an asymmetric potential to obtain the orbital characteristics and the rates of destruction of GCs due to passage through the disk and bulge.

CONSOLIDATED CATALOG OF GLOBULAR CLUSTERS

Prior to starting to perform the integration of GC orbits, we evaluated the observational data given in the published catalogs [6, 32] according to Gaia Data Release 2 (DR2). The catalogs contain the following 6D spatial-phase data for each of the 152 globular clusters: right ascension RA, declination DEC, and heliocentric distance $D_\odot$; proper motions with right ascension (PMRA) and declination (PMDEC); and radial velocity VR. By focusing the error analysis on the proper motions of the GC, we rejected those objects that had a relative error of more than 30% for the radial velocity and the proper motions from further integration [10]. To convert the positions and velocities obtained from the above catalogs into the Galactocentric reference system [16], we took the distance of the Sun from the GalC as $X_\odot$ = 8.178 kpc [14] and $Z_\odot$ = 20.8 pc, the velocity of the local standard of rest (LSR) as $V_{LSR}$ = 234.737 km/s [20], the peculiar speed of the Sun relative to the LSR as $U_\odot$ = 11.1 km/s, $V_\odot$ = 12.24 km/s, and $W_\odot$ = 7.25 km/s [31]. Accordingly, we obtained a consolidated catalog of 143 globular clusters for further study.

To check the possible influence of measurement errors and, first of all, the influence of the errors of proper motions and radial velocities on the obtained results, we generated 1000 random implementations of the initial conditions for each GC. The positions (RA, DEC, and $D_\odot$) in each implementation were kept constant, and normal distribution functions within ±σ were applied for proper motions PMRA and PMDEC with radial velocity VR. We took error values for the proper motions (ePMRA and ePMDEC) and radial velocity (eVR) from the catalog [32].

We used the ϕ-GPU high-order parallel dynamic N-body computer code for orbital integration, which is based on a fourth-order Hermite integration scheme with a hierarchical scheme of individual block steps [7, 15]. Each GC was taken as one physical particle with the assigned position and velocity of the GC center based on observations, which was integrated up to 10 Gyr back in time with its current own mass

(constant in time) from the catalog [6]. During the integration of the orbits, the gravitational interaction of GCs with each other was also taken into account.

## TIME-VARYING POTENTIALS

To relate the integration of GC orbits to the real structure of the Galaxy, we used data from the IllustrisTNG-100 cosmological simulation database and created external dynamic potentials using the approximation method. The IllustrisTNG-100 database has a characteristic volume of approximately 100 Mpc³ and is the second highest resolution cosmological simulation database from publicly available databases. Thus, IllustrisTNG-100 is large enough in volume to contain many individual galaxies similar to the Milky Way. The mass resolution in TNG-100 is $7.5 \times 10^6$ $M_\odot$ and $1.4 \times 10^6$ $M_\odot$ for dark matter and baryon particles, respectively. Given that galaxies similar to the Milky Way have a dark matter halo of approximately $10^{12}$ $M_\odot$ and a disk of approximately $10^{10}$ $M_\odot$, we identify candidate galaxy models with $N = 10^5$-$10^6$ for dark matter and $N = 10^3$-$10^4$ for baryonic matter. To approximate the composite Galactic potential according to the TNG-100 data, we used the formula in which the Miyamoto‑Nagai disk [21] and the Navarro‑Frenk‑White halo [24] are represented as follows:

$$\Phi_{tot}(R,z) = \Phi_D(R,z) + \Phi_H(R,z)$$

$$= -\frac{GM_D}{\sqrt{R^2 + \left(a_D + \sqrt{z^2 + b_D^2}\right)^2}} - \frac{GM_H \cdot \ln(1 + \frac{\sqrt{R^2 + z^2}}{b_h})}{\sqrt{R^2 + z^2}},$$

where $R$ is the distance in the Galactocentric $X$-$Y$ plane, calculated as $R = \sqrt{x^2 + y^2}$; $z$ is the height component of the disc; $G$ is the gravitational constant; $a_D$ is the characteristic horizontal length of the disc, $b_D$ and $b_H$ are the characteristic vertical lengths of the disc and the halo, respectively; $M_D$ and $M_H = 4\pi\rho_0 b_H^3$ are the masses of the disc and the halo, respectively ($\rho_0$ is the central density of the halo).

To obtain the time-dependent gravity, we interpolated between the parameters of each time slice (snapshot) of 1 Myr. Thus, we selected five potentials that are similar in their characteristics to the current parameters of the Milky Way out of 54 potentials[1] [4, 8, 9]. The following parameters are given in Table 1 for the five selected potentials: the total dynamic masses of the disk and the halo and the characteristic lengths of the disk and the halo. The procedure is described in detail in [20].

**Table 1.** Time-varying potential parameters selected from the IllustrisTNG-100 cosmological database

---
[1] https://sites.google.com/view/mw-type-sub-halos-from-illustr/

| Parameter | Value | #411321 | #441327 | #451323 | #462077 | #474170 |
|---|---|---|---|---|---|---|
| Mass of disk, $M_D$ | $10^{10}\,M_\odot$ | 7.110 | 7.970 | 7.670 | 7.758 | 5.825 |
| Mass of halo, $M_H$ | $10^{12}\,M_\odot$ | 1.190 | 1.020 | 1.024 | 1.028 | 0.898 |
| Characteristic length, $a_D$ | 1 kpc | 2.073 | 2.630 | 2.630 | 1.859 | 1.738 |
| Characteristic length, $b_D$ | 1 kpc | 1.126 | 1.356 | 1.258 | 1.359 | 1.359 |
| Characteristic length, $b_H$ | 10 kpc | 2.848 | 1.981 | 2.035 | 2.356 | 1.858 |

The evolution of the main characteristics of #411321 TNG potential is shown in Fig. 1 as an example. Figure 2 shows the evolution of the circular velocity in the Galactic disk as a function of backward time at the distance from the Sun ($R_\odot \approx 8$ kpc). As can be seen from Figs. 1 and 2, the main parameters of the TNG potential have not changed over the past several Gyr and roughly coincide with the current parameters of our Galaxy.

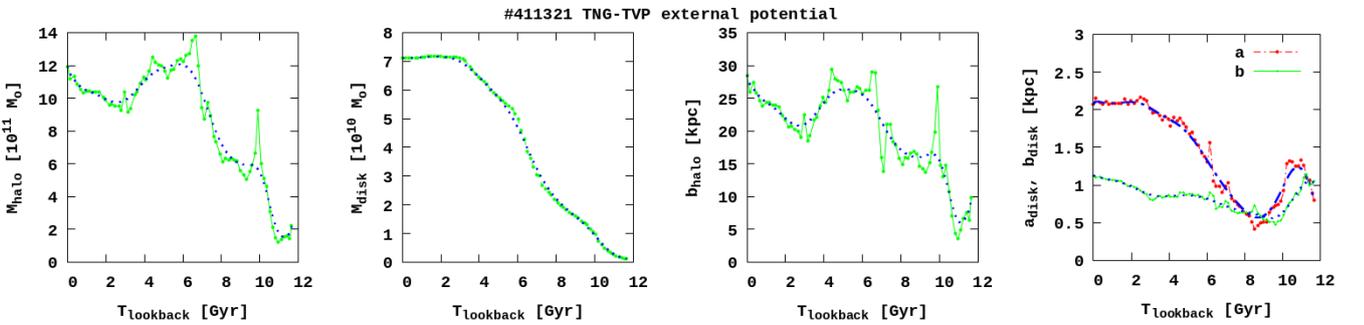

**Fig. 1.** Evolution of the halo and disk masses and their characteristic lengths for TNG potential #411321 (1 Gyr = $10^9$ years). From left to right: halo mass $M_H$, disc mass $M_D$, vertical length parameter $b_H$ of the halo, and characteristic parameters $a_D$ and $b_D$ of the horizontal and vertical lengths of the disc, respectively. The dots show the original data obtained from the IllustrisTNG-100 database, and the dashed lines represent the smoothed approximation values used in the study.

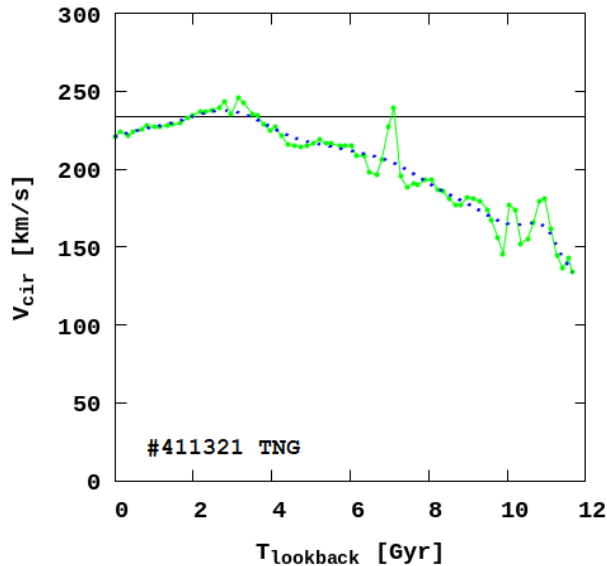

**Fig. 2.** Evolution of the circular velocity as a function of backward time at the distance from the Sun ($R_\odot \approx 8$ kpc) in the Galactic disk. The straight line indicates the current rotation speed of the Sun ($V_\odot \approx 235$ km/s [20]).

## ESTIMATING THE PROBABILITY OF ONE-TO-ONE INTERACTION OF GLOBULAR CLUSTERS

Given the age of such objects in the Galaxy as the GC, it can be predicted that they had a certain percentage of the probability of interacting with each other during their orbital evolution. An exchange of stellar populations during close passages with respect to each other can also be assumed. The first goal of our study was to test this assumption from a statistical point of view. In this study, we used the φ-GPU high-order parallel dynamic N-body computer code [7, 15]. Orbital integration was performed for 143 GC systems from our consolidated catalog that covers 10 Gyr back in time ($T_{back}$). Each GC was integrated as one separate physical particle that has its own mass, the value of which was taken from the catalog [6].

Using 1000 implementations with different initial velocities, we can roughly estimate how likely it is to get a close passage of the GC with respect to each other during the evolution. Such simulations were performed for five external Galactic potentials chosen by us to reproduce the structure of the potential variation in the Galaxy. We consider close passages (cp) to be passages with a relative distance of less than 100 pc and a relative velocity of less than 250 km/s [5]. We believe that the passage of clusters at longer distances and/or with higher velocities does not have a substantial effect on GC orbits. Therefore, such passages are not analyzed in this study.

To analyze the total number of close passages of GCs with respect to each other, we calculated cumulative number $N_{cp}= dN_{cp}/dt$ of GC passages as a function of minimum target parameter $dR_{cp}$, as shown in Fig. 3. As expected, the distribution can be described by a simple power function of the target parameter:

$$\frac{dN_{cp}}{dt}(dR_{cp}) = 10^{a \cdot lg(dR_{cp})+b} \qquad (1)$$

where $N_{cp}$ is the number of close passages of GCs with respect to each other; $a$ and $b$ are the parameters of the power function, which are given in Table 2.

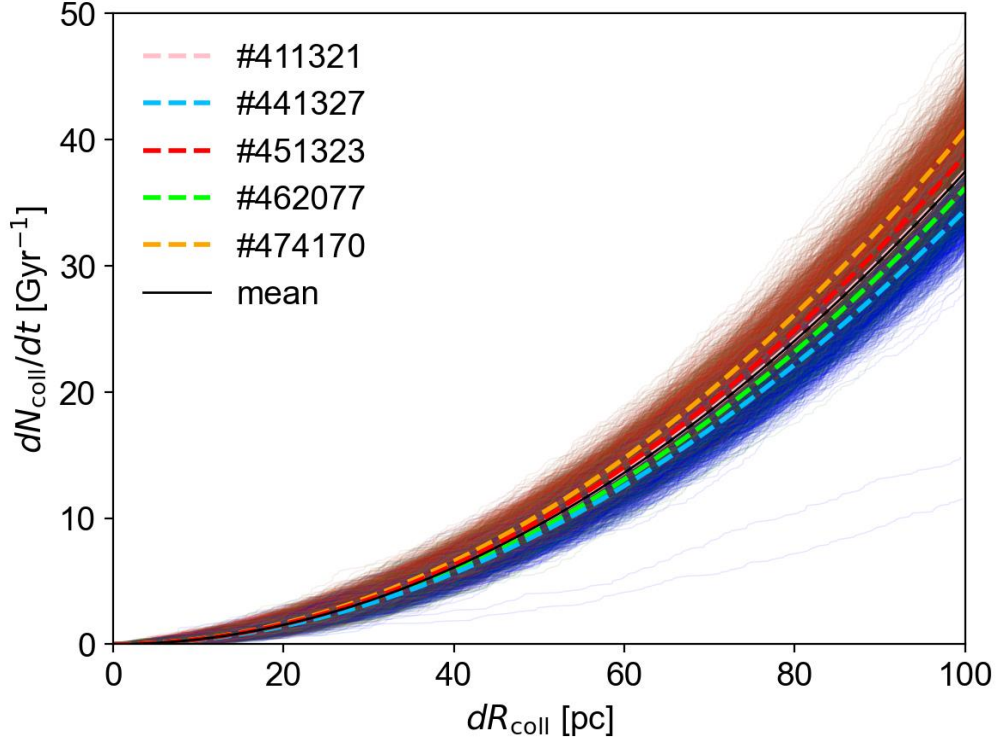

**Fig. 3.** Passages of GCs with respect to each other as a function of the relative distance between them for five potentials and 1000 implementations for each potential (color lines). Lines of different types are means for each of the potentials, which are described by a power law for the relative distance between GCs (see equation (1)). The solid black line is the mean of all calculations for GC systems (see Table 2).

As can be seen from Table 2, the $a$ and $b$ parameters of the power function (equation 1) for all five potentials have a small scatter ($a$ = 1.98 ±0.01 and $b$ = -2.39 ±0.02). That is, the rate of passage is described by simple quadratic formula $dN_{cp}/dt \propto (dR_{cp})^2$, which is a consequence of the formula for the cross-sectional area of an GC that can be crossed by another GC based on general considerations. Analyzing Fig. 3, we can make a general conclusion that we expect approximately 10 GC passages on average at a relative distance of 50 pc in 1 Gyr.

**Table 2.** Parameters *a* and *b* for the rate of interaction of GCs with each other (equation (1)) for five potentials

| Potential | a | b |
|---|---|---|
| #411321 | 1.98 ± 0.16 | -2.38 ± 0.31 |
| #441327 | 1.97 ± 0.17 | -2.41 ± 0.34 |
| #451323 | 1.99 ± 0.11 | -2.38 ± 0.23 |
| #462077 | 1.99 ± 0.15 | -2.43 ± 0.30 |
| #474170 | 1.99 ± 0.14 | -2.37 ± 0.28 |
| Mean | 1.98 ± 0.01 | -2.39 ± 0.02 |

In our opinion, the largest changes in the orbits can be caused by the passages of clusters with low velocities at distances smaller than a few fold (for example, four) the sum of the radii of the half-masses of clusters *i* and *j*. Therefore, we take into further consideration the transitions for which condition $dR_{cp} < 4(R_{hm, i} + R_{hm, j})$ is fulfilled. For the sake of brevity, we will call such passages "collisions". To carry out a visual analysis of the statistical probability of such collisions of GCs with each other, we used the 3D Morton sequence to sort them [19, 23, 30]. Figure 4 gives the statistical probability analysis based on the values of energy *E*, total angular momentum *L*, and the *z*-th component of angular momentum $L_z$ for each GC. The application of this analysis allows one to replace the three-dimensional distribution with a one-dimensional one. Along this one-dimensional distribution, we sort the GC system. The hues of color in Fig. 4 show the statistical probability of collision of GC with each other.

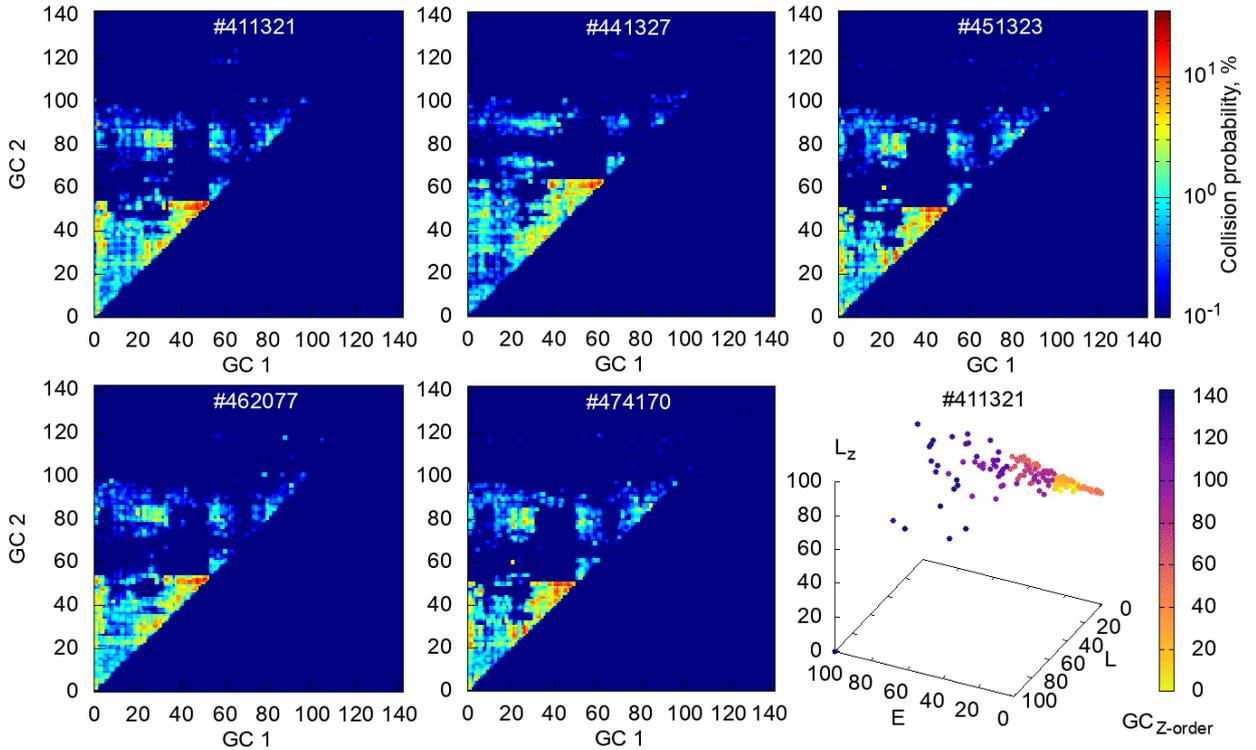

**Fig. 4.** Statistical probability of collision of GCs with each other for five potentials after sorting objects by the Morton analysis. The values of energy $E$, total angular momentum $L$, and the $z$-th component of angular momentum $L_z$ for each GC were used as input data for sorting. The probability percentage of passage is indicated by the hues of color (upper right corner). The $X$ and $Y$ axes show the GC parameters obtained by the Morton analysis. The lower right corner shows the position of individual GCs in the three-dimensional space of normalized $E$-$L$-$L_z$ values for potential #411321. The gray color gradation indicates the order of the GC after the Morton analysis (lower right corner).

As can be seen from the probability distribution given in Fig. 4, certain groups of GCs sorted according to their physical parameters (energy, total angular momentum, and $z$-th component of angular momentum) form a fairly high percentage of collision probability based on the results of 1000 simulations (approximately 30%, shown by red color).

Analyzing the results for 1000 models, we determined the percentage of probability that such an event could occur for each of the five potentials. Table 3 shows an example of GCs that have a high probability of collision with each other for all five potentials. The percentages are calculated for all five potentials and the average value is given. The values of the half-mass radii for the corresponding GCs in pc are also given. Thus, we analyzed the statistical data of the probability of GCs interacting with each other on the basis of 1000 implementations with different initial conditions (variation of velocities along three coordinate components in the Galactocentric reference frame within the measurement error). The GC orbits were integrated over 10 Gyr backward in time with the use of five potentials to realistically reproduce the change in the physical parameters of the Galaxy. Analyzing the function of the relative distance between all GCs selected by two criteria (mutual distances and velocities of GCs with respect to each other), we have approximately ten close passages in 1 Gyr at a distance of less than 50 pc for each of the potentials. We obtained a significant probability of up to 30% for collisions of some GCs with each other when applying three criteria (additional restriction on the mutual distance), as shown in Table 3.

**Table 3.** Example of several GCs that have a high probability percentage of the events of collision with each other for five potentials

| GC1 | $R_{hm1}$ | GC2 | $R_{hm2}$ | #411321 | #441327 | #451323 | #462077 | #474170 | Mean |
|---|---|---|---|---|---|---|---|---|---|
|  | pc |  | pc | % | % | % | % | % | % |
| Terzan 4 | 4.13 | NGC 6440 | 2.20 | 25.0 | 24.0 | 29.1 | 22.8 | 30.0 | 26.2 ± 3.2 |
| Terzan 4 | … | Terzan 5 | 2.22 | 20.0 | 11.0 | 21.2 | 15.4 | 22.4 | 18.0 ± 4.7 |
| Terzan 4 | … | Terzan 9 | 1.83 | 27.5 | 16.3 | 21.4 | 25.7 | 28.3 | 23.8 ± 5.0 |
| Terzan 4 | … | NGC 6624 | 2.57 | 21.5 | 13.0 | 22.4 | 23.9 | 25.2 | 21.2 ± 4.8 |
| Terzan 2 | 3.70 | Terzan 4 | 4.13 | 21.3 | 20.3 | 28.7 | 21.8 | 21.8 | 22.8 ± 3.4 |
| Terzan 2 | … | Terzan 6 | 2.58 | 15.8 | 12.3 | 19.9 | 13.4 | 21.4 | 16.6 ± 4.0 |

## ESTIMATING THE PROBABILITY OF INTERACTION OF GLOBULAR CLUSTERS WITH THE GALACTIC CENTER

The second goal of our study was to determine the probability of the events of close passage of GCs with the GalC. To calculate the number of GCs crossings with the GalC, we applied the criterion of the minimum distance of the GC to the GalC, which should be <100 pc. We also used the ϕ-GPU high-order parallel dynamic N-body computer code for solving this problem. Orbital integration was also performed for 143 GC systems over 10 Gyr backward in time.

Using 1000 implementations with initial conditions, we can roughly estimate how likely it is to get a close passage of the GC with the GalC during its evolution. Such simulations were performed for the five outer galactic potentials we had selected earlier. Having analyzed the results of the obtained GC passes near the GalC for 1000 implementations, we determined the average probability percentage at which such events could occur. Table 4 shows a list of GCs that have a different percentage of the probability of close passages with the GalC for all five potentials. The percentages were calculated for each of the GCs for all potentials, and the average value for each of the GCs was determined.

**Table 4.** Example of several GCs with different parameters of the probability percentage of the event of passing near the GalC for five potentials in %

| GC | #411321 | #441327 | #451323 | #462077 | #474170 | Mean |
|---|---|---|---|---|---|---|
| NGC 1904 | 28.1 | 29.8 | 34.1 | 30.1 | 29.7 | 30.4 ± 2.2 |
| Pal 6 | 0.7 | 0.6 | 25.1 | 12.4 | 0.8 | 7.9 ± 10.9 |
| NGC 6642 | 64.3 | 30.3 | 96.7 | 85.1 | 95.9 | 75.6 ± 27.9 |
| NGC 6981 | 45.8 | 43.6 | 54.0 | 41.9 | 52.6 | 47.6 ± 5.4 |

Figure 5 shows examples of the evolution of the orbits of the NGC 1904, Pal 6, NGC 6642, and NGC 6981 clusters, which satisfy the requirement of a close passage near the GalC up to 100 pc for potential #411321. The orbits are given in three coordinate planes $X$-$Y$, $X$-$Z$, and $R$-$Z$, where $R$ is the distance in the Galactocentric $X$-$Y$ plane and calculated as $R = \sqrt{x^2 + y^2}$. One can see that the GC orbits evolve over time. As the total mass of the disk and the halo of the Galaxy decreases, the GC orbits move to larger distances over time.

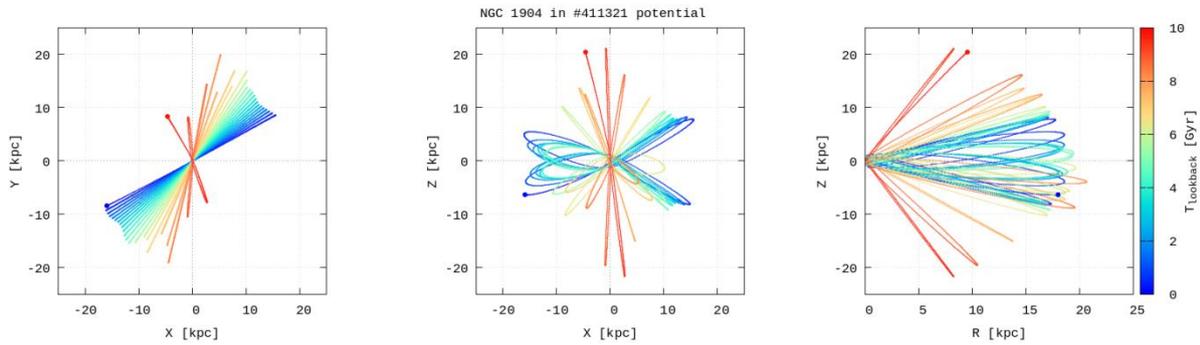

(a) NGC 1904

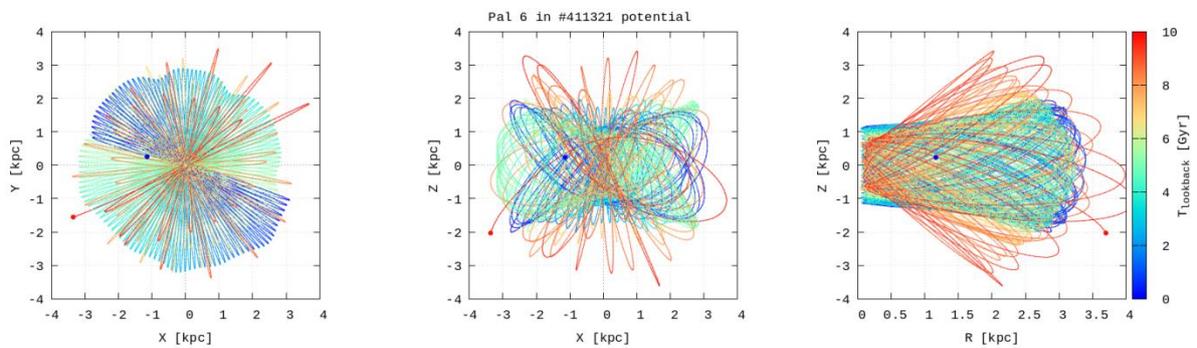

(b) Pal 6

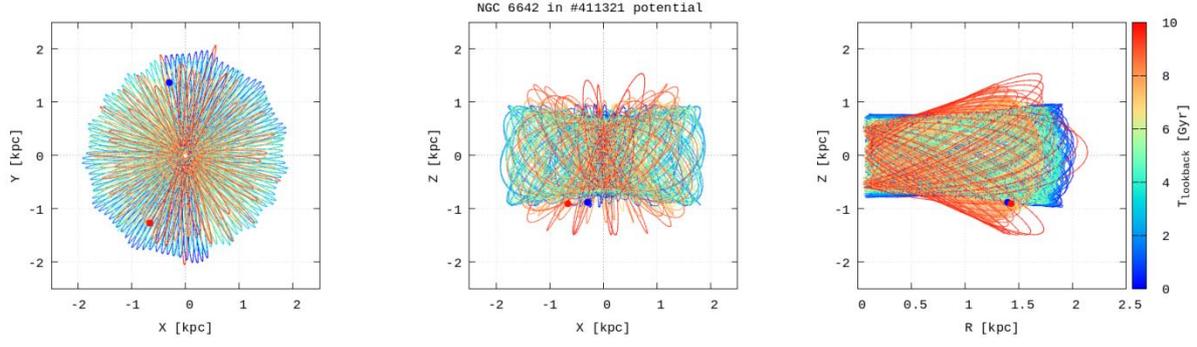

(c) NGC 6642

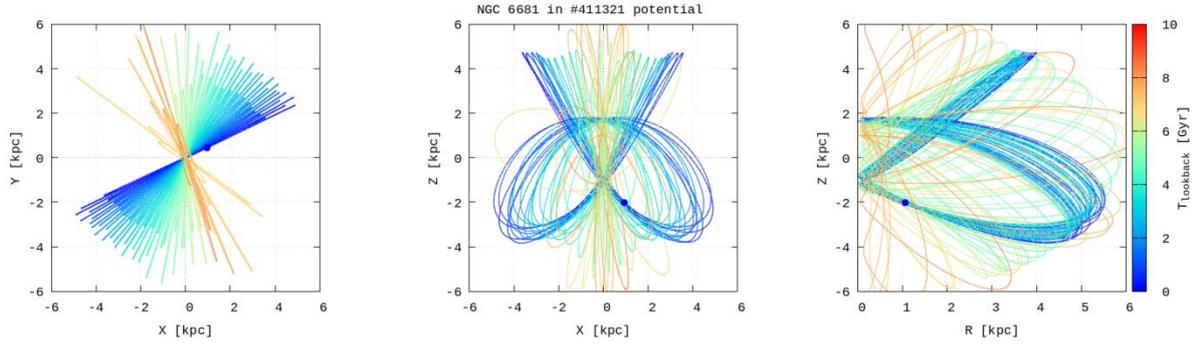

(d) NGC 6981

**Fig. 5.** Examples of the evolution of the orbits of GCs that have close passages near the GalC at a distance of up to 100 pc for potential #411321: (a) NGC 1904, (b) Pal 6, (c) NGC 6642, (d) NGC 6981. The time of integration (up to 10 Gyr backward in time) is indicated by the hue of color. The blue and red points in the diagrams are the initial and final coordinates of the GC during time integration.

Figure 6 shows an example of changing the distance from the GalC for some GCs under the influence of the physical difference in potential parameters, for example, potentials #411321 and #441231. The galactocentric distance $D_G$ was calculated using three coordinate components, i.e., $D_G = \sqrt{x^2 + y^2 + z^2}$. An example is given for the NGC 1904, Pal 6, NGC 6642, and NGC 6981 globular clusters.

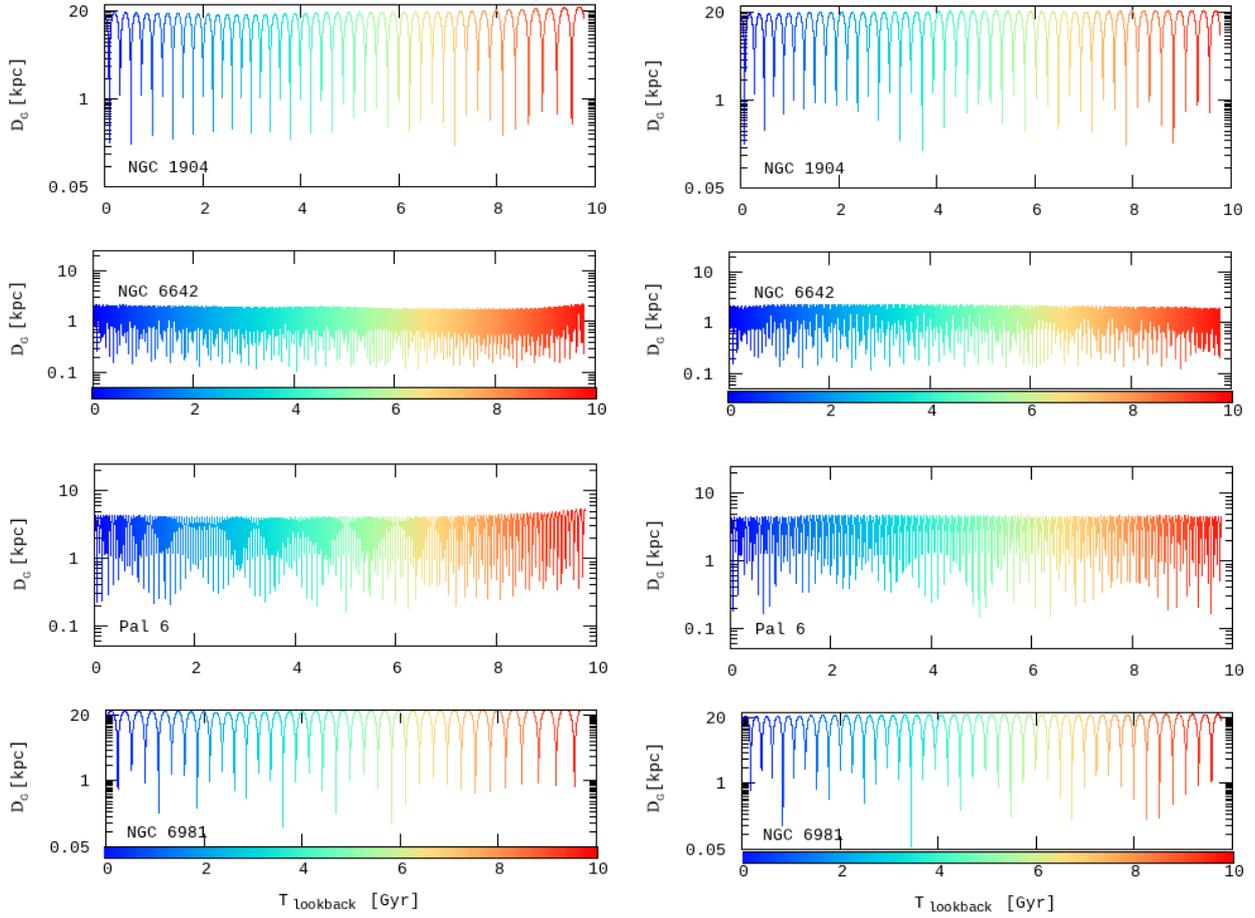

**Fig. 6.** Variation of Galactocentric distance $D_G = \sqrt{x^2 + y^2 + z^2}$ from the GalC for globular clusters NGC 1904, Pal 6, NGC 6642, and NGC 6981 (from top to bottom) during the orbital evolution for potentials (a) #411321 and (b) #441231. The color gradation marks the integration time backward down to 10 Gyr ago.

As can be seen from Fig. 6, close passages occur almost during the entire orbital evolution of the GC. It should be noted that is not always possible to see the passage of the GC beyond the limit of 100 pc in the graph. This is determined by the fact that these graphs were constructed using data that are derived from the calculations with a relatively large fixed time step of 1.22 Myr. At the same time, the GC passage up to 100 pc is recorded directly during the integration of the orbit in the computer code itself by the built-in analyzer. Comparing the deviation of the Galactocentric distance for the two potentials, one can see that there are small changes in the distance caused by different values of the energy of the potentials.

We also evaluated cumulative value *dN/dt* of the interaction of GCs with the GalC, which is given as a function of the minimum value of target parameter $D_G$ (Fig. 7). We tried to describe the distribution as a function of the target parameter by a simple power function (see equation (1)), for which the *a* and *b* parameters are given in Table 5.

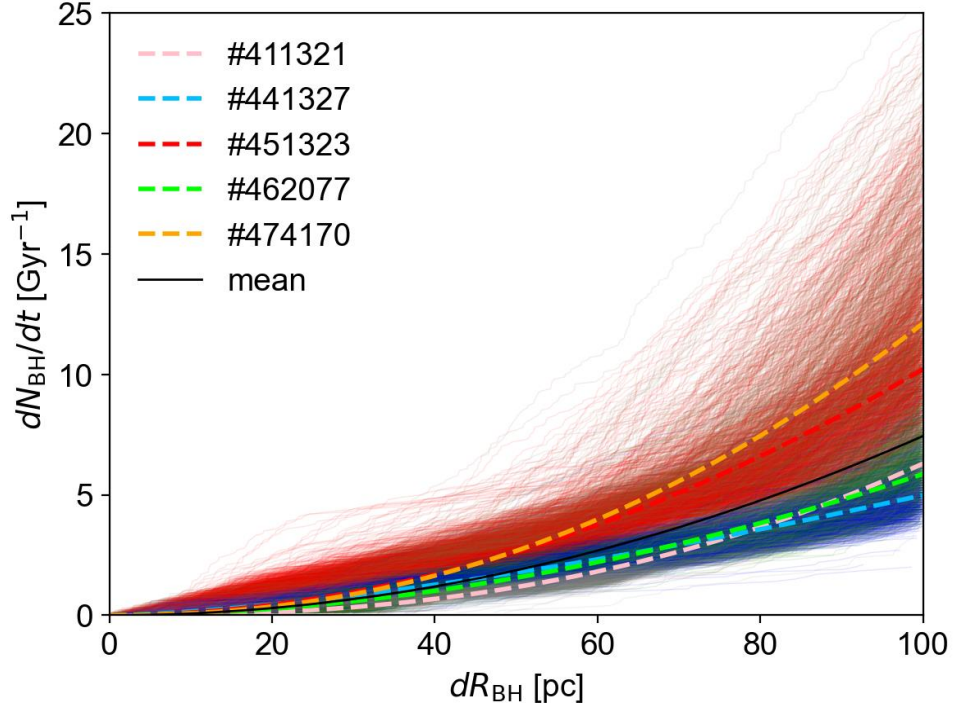

**Fig. 7.** Interaction of the GC with the GalC as a function of the relative distance to the GalC for five potential models and 1000 implementations for each of them (solid gray lines). Lines of different types are power-law functions for the relative distance between the GC and the GalC at the corresponding potentials (see equation (1)). The solid black line is the averaged function for all simulations (see Table 5).

**Table 5.** Parameters *a* and *b* for the rate of interaction of the GC with the GalC (equation (1)) for five potentials

| Potential | *a* | *b* |
|---|---|---|
| #411321 | 2.43 ± 0.67 | -4.06 ± 1.26 |
| #441327 | 1.91 ± 0.53 | -3.06 ± 1.00 |
| #451323 | 2.18 ± 0.55 | -3.28 ± 1.01 |
| #462077 | 1.96 ± 0.62 | -2.90 ± 1.16 |
| #474170 | 1.50 ± 0.35 | -2.29 ± 0.65 |
| Mean | 2.00 ± 0.35 | -3.12 ± 0.64 |

As can be seen from Table 5, mean parameters *a* and *b* of the power function (see equation (2)) for all five models of potentials are $a = 2.00 \pm 0.35$ and $b = -3.12 \pm 0.64$. Hence, the rate of close passages is also roughly described by the simple quadratic formula $dN_{cp}/dt \propto (D_G)^2$, but we see significant deviations from a simple quadratic dependence compared to the rate given in Table 2.

A general conclusion can be also drawn analyzing the data given in Fig. 5; for example, it can be concluded that approximately three to four close passages in 1 Gyr occur at a distance of 50 pc between the GC and the GalC. Thus, we analyzed the statistical data of the probability of close passage of the GC near the GalC obtained from 1000 implementations with different initial conditions. The orbits of 143 GCs were integrated over 10 Gyr backward in time with use of five potentials.

Thus, we analyzed the statistical data of the probability of close passage of the GC near the GalC obtained from 1000 implementations with different initial conditions. The orbits of 143 GCs were integrated over 10 Gyr backward in time with use of five potentials. By applying the criterion of close passage of GCs near the GalC up to 100 pc, we obtained a high probability of such phenomena for some GCs. Having analyzed the function of the relative distance between the GC and the GalC, we revealed that approximately three to four close passages per 1 Gyr occur at a distance of up to 50 pc for each of the potentials.

## CONCLUSIONS

The dynamic evolution of GC orbits 10 Gyr backward in time are investigated using timevarying potentials that are closest to the physical parameters of the potential of our Galaxy. The orbital trajectories for 143 globular clusters are integrated using our original ϕ-GPU high-order N-body parallel dynamic computer code. For each of the potentials, 1000 input files with randomized initial velocities within the errors of the observational data are generated for the GCs.

To detect clusters that pass close to each other, we use the following two criteria: (1) the distance between GCs should be less than 100 pc; (2) the value of the relative velocity between GCs should not be greater than 250 km/s. To identify GCs that have the potential to collide during their evolution, we add criterion (3) that ensures that the distance between GCs should be less than fourfold of the sums of the half-mass radii of these clusters.

To detect clusters that have passages close to the GalC, we apply the criterion according to which the relative distance should be less than 100 pc.

The obtained results answer the question about the possibility of events and also allow one to estimate the probability of close passages and collision of GCs with each other and with the central supermassive black hole in the past. Applying the above criteria, we have obtained statistically significant rates of close passages of GCs with respect to each other and to the GalC at a distance of less than 100 pc. We have determined that GCs have approximately ten passages at a close distance with each other during their evolution and approximately three to four passages near the GalC in 1 Gyr at a distance of up to 50 pc for each of the potentials. Moreover, certain groups of GCs sorted according to physical parameters (energy, total angular momentum, and the

*z*-th component of angular momentum) are characterized by a fairly high percentage of the probability of collision based on the results of 1000 simulations.

## ACKNOWLEDGMENTS

We thank the reviewer for his time and valuable comments that improved the presentation.

## FUNDING

The work of P. Bertsyk and T. Panamarev were partially supported by the Scientific Committee of the Ministry of Education and Science of the Republic of Kazakhstan (grant AP AP08856184).

The research activities of M. Ishchenko and M. Sobolenko were supported by the National Academy of Sciences of Ukraine under research project no. 0121U111799 for young scientists.

The research activities of P. Berczik and M. Ishchenko were partially supported by joint grant no. M/32-23.05.2022 from the Ministry of Education and Science of Ukraine.

The research activities of P. Berczik, M. Ishchenko, and M. Sobolenko were partially supported by Volkswagen Foundation within the framework of partnership grant nos. 97778.

The research activities of P. Berczik and M. Ishchenko were partially supported by project no. 13.2021.MM of the National Academy of Sciences of Ukraine for development of the GPU computing cluster.

M. Sobolenko is grateful for a scholarship of the National Academy of Sciences of Ukraine in 2020–2022 and 2022-2024.

P. Berczik and M. Ishchenko are grateful for a visiting grant from the Nicolaus Copernicus Astronomical Center of the Polish Academy of Sciences, in which part of the study was performed.

This study used data from the GAIA mission of the European Space Agency (ESA) (https://www.cosmos.esa.int/gaia) processed by the GAIA Data Processing and Analysis Consortium (DPAC, https://www.cosmos.esa.int/web/gaia/dpac/consortium). Funding for DPAC was provided by national institutions, particularly institutions involved in the GAIA Multilateral Agreement.

## CONFLICT OF INTEREST

The authors declare that they have no conflicts of interest.